\documentstyle[aps,prb,12pt]{revtex}

\begin{document}
\title{Effect of Weak Disorder in the Fully Frustrated XY Model.}
\author{Vittorio Cataudella}
\address{I.N.F.M. Unit\'a di Napoli, Dip. di Scienze Fisiche\\
Mostra D' Oltremare pad.19 - 80125 Napoli, Italy}
\maketitle

\begin{abstract}
The critical behaviour of the Fully Frustrated XY model in presence of weak
positional disorder is studied in a square lattice by Monte Carlo methods.
The critical exponent associated to the divergence of the chiral correlation
length is found to be $\nu \simeq 1.7$ already at very small values of
disorder. Furthermore the helicity modulus jump is found larger than the
universal value expected in the XY model.
\end{abstract}


The study of the fully frustrated XY (FFXY) model\cite{villain,halsey} in a
planar lattices has attracted much attention since it provides a simple
model for many physical systems. In particular it is related to
Josephson-junction arrays and superconducting wire networks in a magnetic
field \cite{review} for which recent experiments show interesting properties%
\cite{ling}. In these systems a perpendicular magnetic field $B$ penetrates
the array inducing a quenched vortex density. When the field is such that
each plaquette of area $S$ in the planar lattice traps half a vortex, e.g.
the ratio of the magnetic field to the flux unit $f=BS/\phi _0=1/2$, the
system can be described by the FFXY model. As first noted by Villain\cite
{villain} the FFXY model posses, besides the continous $U(1)$ symmetry, a
discrete $Z_2$ symmetry which is not present in the unfrutrated XY model.
The existence of two different symmetries leads to the possibility of two
critical temperatures that have been investigated by both analytical methods%
\cite{halsey,night} and Monte Carlo (MC) simulations\cite
{teitel,granato1,jose,olssonf1,grest}. However, in spite of accurate MC
simulations, the scenario is not fully understood. In fact, there are
numerical evidences supporting both the existence of two very close critical
temperatures ($T_c^{Z_2}>T_c^{U(1)}$) with critical behavior typical of
Ising and Thouless-Kosterlitz-Berezinski (TKB) transitions respectively\cite
{olssonf1,grest,olssonf2} and the existence of a single transition with
novel critical behavior\cite{granato1}. In the latter case it has been
claimed that the transition associated to the $Z_2$ symmetry is
characterized by a correlation length critical exponent $\nu $ lower than
that expected in the Ising case.

Recent experiments on Nb wire networks\cite{ling} in an external magnetic
field have shown that the critical exponent $\nu $ which can be extracted by
the scaling of current-voltage (I-V) characteristics, is in agreament
neither with a Ising-like exponent ($\nu =1$) nor with the value suggested
in Ref.[\cite{granato1} ] ($\nu \simeq 0.83$). On the contrary, the scaling
of the experimental data suggests a much larger value.

The aim of the present work is to investigate if the effect of weak quenched
disorder in the FFXY model is able to change the critical behaviour of the
system and to explain the unexpected critical behaviour.

The hamiltonian of the FFXY model is

\begin{equation}
H=-J\sum_{\langle ij\rangle }cos(\theta _i-\theta _j-A_{i,j}),
\label{hamiltonian}
\end{equation}
where $\theta _i$ is the phase in the site $i$ and 
\begin{equation}
A_{i,j}=\frac{2\pi }{\phi _0}\int_i^j\vec{A}\cdot d\vec{l}
\end{equation}
is the line integral of the vector potential along the line ($i-j$) and $%
\phi _0$ is the flux quantum. In order to simulate positional disorder we
consider a uniform distribution of the $A_{i,j}$ whose average value $%
\langle A_{i,j}\rangle $ is equal to $\pi $ every second column and zero
otherwise. The disorder configuration is thus specified by the probability
distribution

\begin{equation}
p(A_{i,j})=\left\{ 
\begin{array}{lll}
\frac 1{2\phi _m} &  & -\phi _m<(A_{i,j-}\langle A_{i,j}\rangle )<\phi _m \\ 
0 &  & \text{otherwise}
\end{array}
\right.
\end{equation}
The parameter $\phi _m$ controls the degree of randomness and is related to
the variance according to $\sigma ^2=\phi _m^2/3$.

Of course, the presence of positional disorder effects the behaviour of the
model. In fact it has been shown by renormalization group techniques\cite
{granatod} and Monte Carlo simulations\cite{choi} that the critical
temperatures associated to both $Z_2$ and $U(1)$ simmetries decrease with
increasing disorder. It has been also suggested for the $U(1)$ transition a
reentrant behaviour, even if recent work on the disordered unfrustrated XY
model seems to exclude this possibility\cite{scheidl}. However, the critical
behaviour that characterize the Ising-like and the TKB transitions have not
been studied to our knowledge. In particular, the study of the critical
exponent $\nu $ associated to the divergence of the coherence lenght of the
Ising-like transitions is interesting in view of its possible connection
with the $I-V$ characteristics reported by Ling\cite{ling}. In fact the
experimental data I-V, or equivalently the data J-E, scale very well
according to the relation\cite{fisher} 
\begin{equation}
E=J\mid T-T_c\mid ^{\nu z}G(J\mid T-T_c\mid ^{-\nu })
\end{equation}
where the dynamical critical exponent $z$ is found to be about 2 and $\nu
=1.7\pm 0.5$. This type of scaling is consistent with the reasonable
assumption that the relevant excitations that contribute to the critical
behaviour of the I-V characteristic are domain-wall.

In order to obtain an estimation of $\nu $ we have calculated the Binder's
cumulant of the staggered magnetization $M$ 
\begin{equation}
U=1-\frac{\langle M^4\rangle }{3\langle M^2\rangle ^2}.  \label{binder}
\end{equation}
where $M$, associated to the chirality of each plaquette, is defined as 
\begin{equation}
M=\frac 1{L^2}\mid \sum_{\vec{r}}(-1)^{r_x+r_y}m_{\vec{r}}\mid .  \label{m}
\end{equation}
In eq. (\ref{m}) the chirality $m_r$ is the rotation of the current around a
plaquette 
\begin{equation}
m=\frac 1{\sqrt{8}}(sin\phi _{12}+sin\phi _{23}+sin\phi _{34}+sin\phi _{41})
\end{equation}
where $\phi _{ij}$ is the phase difference at the edges of the plaquette.

Binder's cumulant allows an accurate estimation of the critical temperature
associated to the ordering of the plaquette chirality (Ising-like
transition). In fact $U(T_c,L)$ does not depend on lattice size $L$ for
large systems and, then, $T_c$ can be identified without making any
assumption on the critical exponents. Once a satisfactory estimation of $T_c$
is obtained the critical exponent $\nu $ is estimated through a data
collapsing with the parameter $\nu $ only. The Binder's cumulant $U$ has
been calculated by Monte Carlo (MC) simulations using a standard Metropolis
dynamics. For each disorder configuration the mean value $\langle M\rangle $
has been obtained averaging over $4\times 10^5$ MC sweeps after discarding
the first $10^5$ for equilibrating the system. Then the mean values are
further averaged over a number of disorder configurations that range from $%
200$ to $400$ depending on lattice size and amount of disorder present in
the system.

In Fig. 1) it is shown how the critical temperature $T_c^{Z_2}$ can be
extracted by using the MC estimations of $U$ for different lattice sizes. In
the case reported in the figure, if we exclude the smallest lattice size, $%
L=12$, all the data cross about $T_c^{Z_2}=0.438\pm 0.003$. From this type
of analysis it is, then, possible to follow $T_c^{Z_2}$ as a function of the
disorder (Fig.2)). The results confirm that $T_c^{Z_2}$ decrease with
increasing disorder going from $T_c^{Z_2}=0.452\pm 0.002$ for the case
without disorder ($\phi _m=0$) to $T_c^{Z_2}=0.350\pm 0.005$ for the case
with largest disorder studied ($\phi _m=0.245$). The results are in
qualitative agreement with those reported by Choi et al.\cite{choi}. For
larger values of the disorder preliminar results seems to suggest a
continous collaps of $T_c^{Z_2}$. In Figs. 3a) and 3b) we show Binder's
cumulant $U$ as a function of $\mid T-T_c^{Z_2}\mid L^{1/\nu }$ for two
values of $\phi _m$. By using the estimated values of $T_c^{Z_2}$ at $\phi
_m=0.098$ and $\phi _m=0.196$ we see that the best data collapsing is
obtained for $\nu =0.9\pm 0.2$ in the case of lower disorder and for $\nu
=1.7\pm 0.2$ in the other case. Even if the numerical errors do not allow a
very accurate estimation of $\nu $ it is very clear that already for very
weak disorder ($\phi _m>0.19$) the critical exponent is quite different from
that expected in the Ising model ($\nu =1$) and from that suggested by
Granato et al.\cite{granato1} in the case without disorder. Instead the
value obtained is very close to that extracted from the $I-V$ characteristics%
\cite{ling}. It is also interesting to note that a very similar value is
obtained in the 2d XY spin glass that has a critical temperature $T_c^{Z_2}=0
$.\cite{bokil,kawamura,ray} The same analysis for $\phi _m=0$ and $\phi
_m=0.245$ gives clearly $\nu \simeq 1$ and $\nu \simeq 1.7$, respectively.
On the other hand for the intermediate value $\phi _m=0.147$ the data are
not able to clearly exclude either $\nu \simeq 1$ or $\nu \simeq 1.7$;
larger lattices and more MC sweeps are needed to reach a satisfactory
estimation of $\nu $. These results indicate the existence of a threshold
disorder above which $\nu $ changes from 1 to 1.7. The value of the
threshold can be, then, estimated to be $\phi _m\simeq 0.19$ but it must be
understood as an upper limit.

As in the FFXY model we expect the system to present a jump in the helicity
modulus at temperature lower or equal to $T_c^{Z_2}$. In order to study this
possibility we have calculated the helicity modulus that for the system
studied takes the form 
\begin{eqnarray}
\Gamma &=&\frac J{L^2}\left\{ -\frac 1{k_BT}\langle \left( \sum_{\langle
ij\rangle }\sin (\theta _i-\theta _j-A_{ij})\hat{e}_{ij}\cdot \hat{x}\right)
^2\rangle \right.  \nonumber \\
&+&\langle \sum_{\langle ij\rangle }\cos (\theta _i-\theta _j-A_{ij})\left( 
\hat{e}_{ij}\cdot \hat{x}\right) ^2\rangle  \nonumber \\
&+&\left. \frac 1{k_BT}\langle \sum_{\langle ij\rangle }\sin (\theta
_i-\theta _j-A_{ij})\hat{e}_{ij}\cdot \hat{x}\rangle ^2\right\}
\label{gamma}
\end{eqnarray}
where $\hat{e}_{ij}$ and $\hat{x}$ are the unitary vectors in the directions
of the link $i\mapsto j$ and $\vec{x}$, respectively. The symbol $\langle
\cdot \rangle $ indicates the average over the spin configurations. The
elicity modulus $\Gamma $ has to be averaged further over the disorder
configurations to obtain the mean value $\overline{\Gamma }$. Following the
analysis developed for the XY model\cite{weber} and extended to the FFXY
model\cite{olssonf1} we can estimate the critical temperature $T_c^{U(1)}$
by using the following ansatz for the size dependence of $\overline{\Gamma }$

\begin{equation}
\frac{\pi \overline{\Gamma }}{2T_c}=\overline{\Gamma }_0+\frac 1{2(\ln L+\ln
l_0)}  \label{scaling}
\end{equation}
where $\overline{\Gamma _0}=1$ corresponds to the universal jump expected at
the critical temperature in a KTB transition. This critical scaling, being
based on the mapping between a neutral Coulomb gas and a XY model, is valid
only if the transition is of KTB type. Therefore eq. (\ref{scaling}) can be
used as a test for the existence of a standard KTB transition.

In Figs. 4a) and 4b) we show the results of this analysis for two different
values of the disorder. In both cases it is not possible to obtain a
satisfactory fit of the data by using eq. (\ref{scaling}) with $\overline{%
\Gamma }_0=1$. However, a more accurate data analysis shows that a
satisfying fitting of the data can be obtained assuming a helicity modulus
jump larger than the universal value. From such an analysis we get $%
T_c^{U(1)}=0.43\pm 0.003$ and $\overline{\Gamma }_0=1.2$ for $\phi _m=0.098$
and $T_c^{U(1)}=0.38\pm 0.003$ and $\overline{\Gamma }_0=1.7$ for $\phi
_m=0.196.$ In both cases the fitting parameter $l_{0\text{ }}$takes the same
value found in the FFXY model without disorder, $l_0=0.27$,\cite
{olssonf1,vit} and $T_c^{U(1)}$ is slightly smaller than $T_c^{Z_2}$ even if
the estimated numerical error does not rule out $T_c^{U(1)}=T_c^{Z_2}$ in
the case of lowest disorder ($\phi _m=0.098$).\cite{nota} This analysis
shows that the only way to fit the data with eq.(\ref{scaling}) is to assume
a larger helicity modulus jump and this is an indication that the transition
associated to this jump is not a KTB transition.

Summarizing we have calculated, by MC simulations, the critical temperature
and the critical exponent $\nu $ associated to the chiral degree of freedom
in a disordered FFXY model. Unexpected we find that the critical exponent $%
\nu $ already for very weak disorder becomes equal to the value found in the
XY spin glass, i.e. $\nu \simeq 1.7.$ Correspondently, the helicity modulus
jump becomes larger than the universal value expected in KTB transition.
Finally, we note that the value obtained for $\nu $ is in agreement with
recent results in Josephson wire networks.\cite{ling}

\newpage

\section*{Figure captions}

\begin{description}
\item  {Fig. 1 The Binder' s cumulant }$U$ {(see eq.(\ref{binder})) vs.
temperature }$T$ for different sizes $L$. At the critical temperature $U$
does not depend on $L$ for large $L$.

\item  {Fig. 2 The critical temperature }$T_c^{Z_2}$ as a function of the
parameter $\phi _m$ that controls the degree of disorder present in the
system. The continous cirve is a guide for the eye only.

\item  {Fig. 3 The Binder's cumulant (see eq.(\ref{binder})) as a function
of }$(T-T_c^{Z_2})L^{(1/\nu )}$ {for two values of the parameter that
controls the disorder }$\phi _m${. In Fig.3a) the data for }$\phi _m=0.098$
are reported having choosen $T_c^{Z_2}=0.438$ and $\nu =1$. {In Fig.3b) the
data for }$\phi _m=0.196$ are reported having choosen $T_c^{Z_2}=0.395$ and $%
\nu =1.7$. Temperature is in units of $k_B/J$.

\item  {Fig. 4 The scaled helicity modulus }$\frac{\pi \overline{\Gamma }}{%
2T_c}$ as a function of $\ln L$ for two {values of the parameter that
controls the disorder }$\phi _m$. {In Fig.4a) the data for }$\phi _m=0.098$
and $T=0.45,0.44,0.43,0.42$ (from below to the top) are reported. The best
fit is given by the full curve that corresponds to eq.(\ref{scaling}) with $%
\overline{\Gamma }_0=1.2$ and $l_0=0.27$. {In Fig.4b) the data for }$\phi
_m=0.196$ and $T=41,39,38,37$ (from below to the top) are reported. The best
fit is given by the full curve that corresponds to eq.(\ref{scaling}) with $%
\overline{\Gamma }_0=1.7$ and $l_0=0.27$.
\end{description}

\newpage


\begin{references}
\bibitem{nota}  The estimated values of $T_c^{U(1)}$ and $T_c^{TKB}$ show
that increasing the degree of disorder in the system the difference betwwen
the two critical temperatures increases. This is in agreement with the
results reported by Granato and Kosterlitz \cite{granatod}
\end{references}
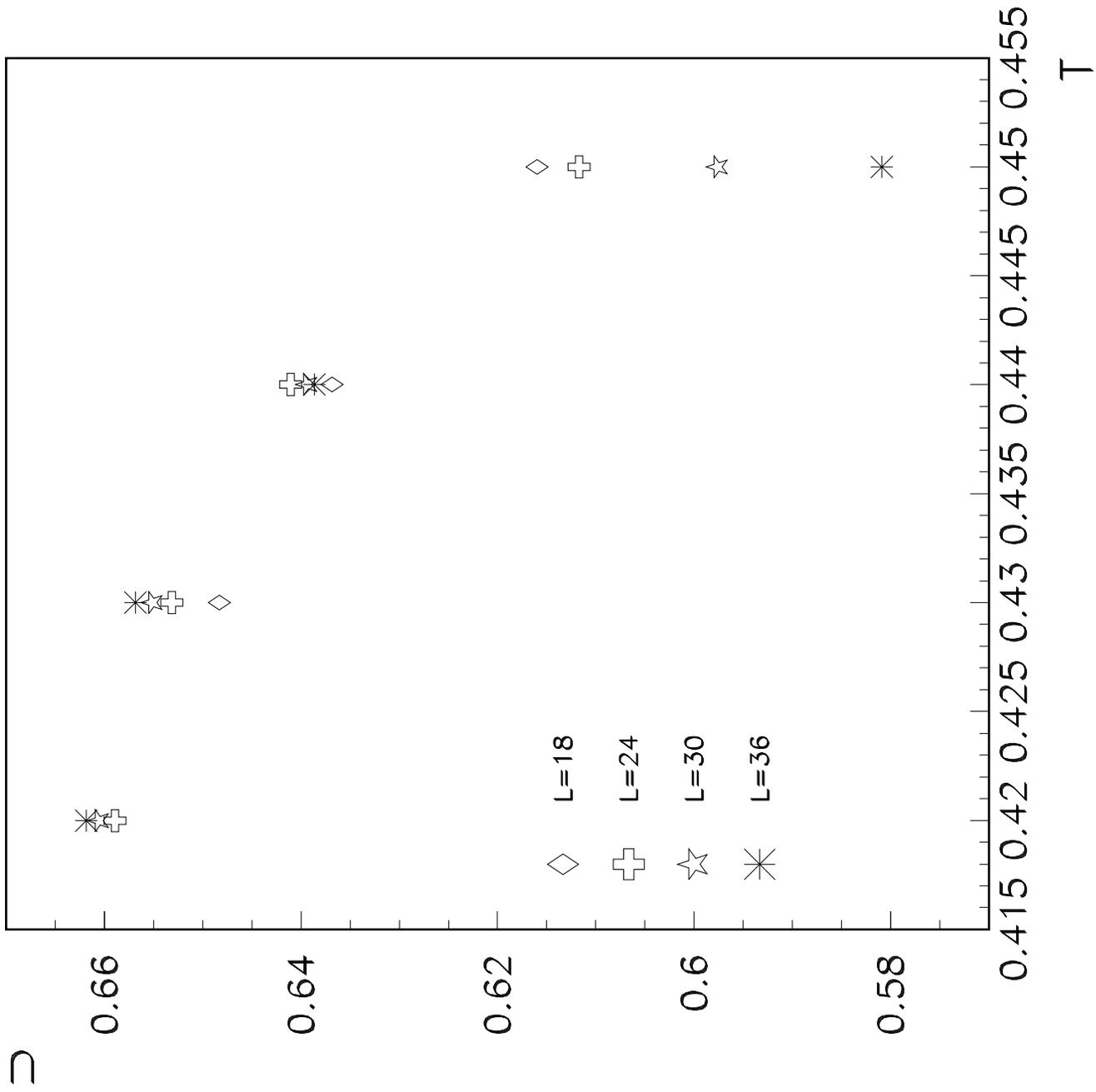
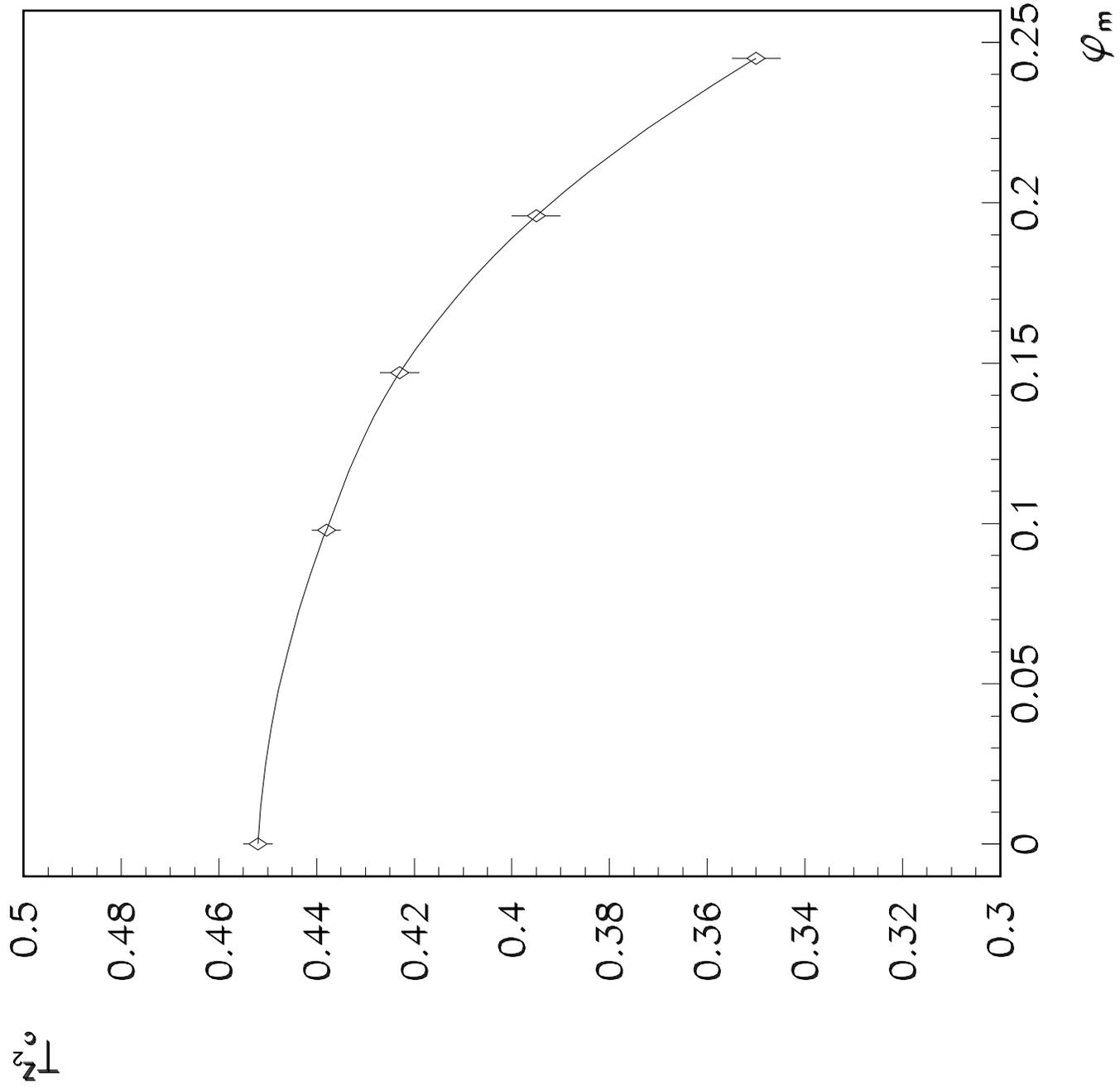
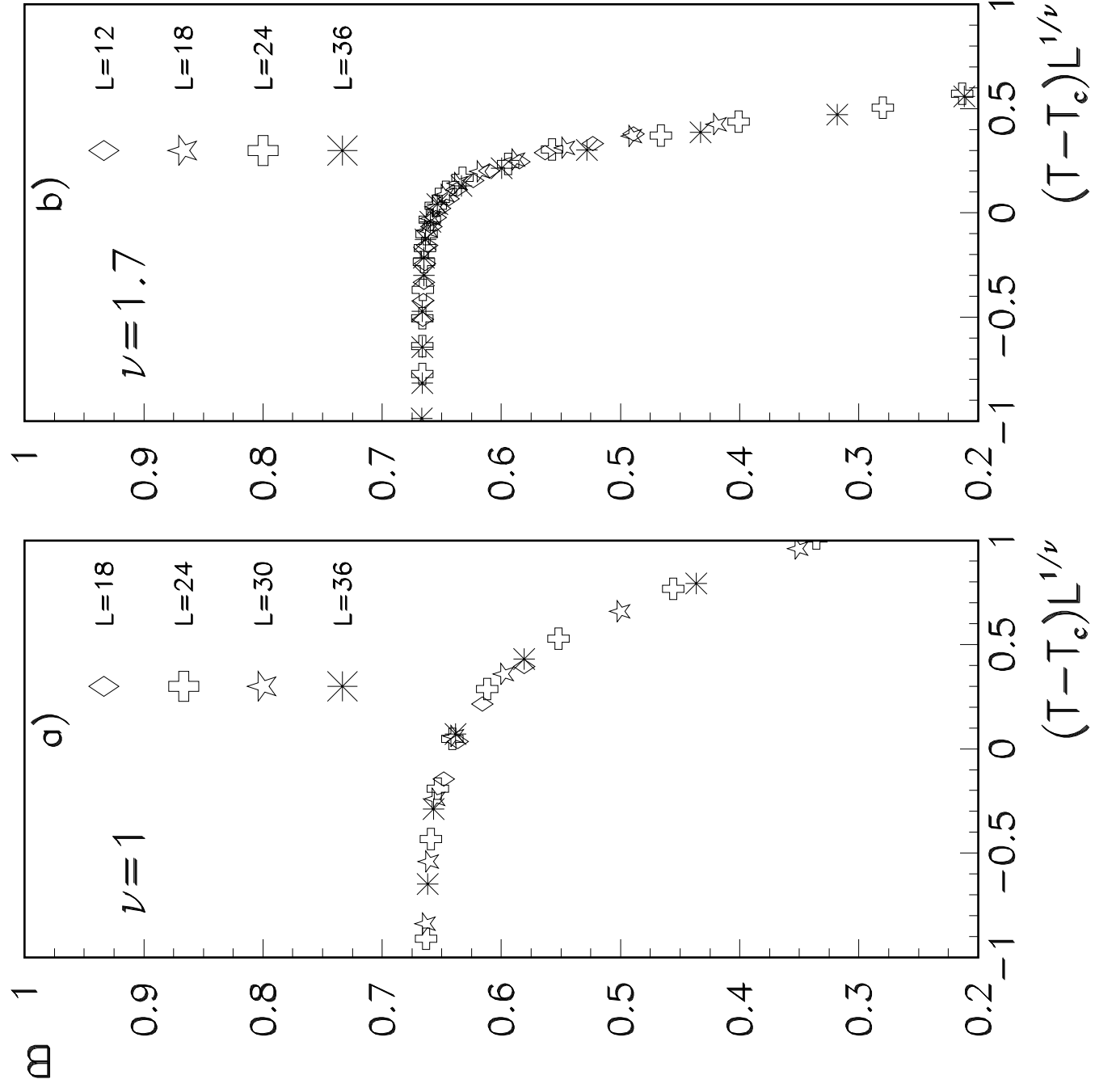
\end{document}